\documentclass[twocolumn,aps,prb,superscriptaddress]{revtex4-1}

\usepackage{amsmath}
\usepackage{amssymb}
\usepackage{graphicx}
\usepackage{epstopdf}
\usepackage[colorlinks=true]{hyperref}

\newcommand{\sech}{\mathrm{sech}}

\begin{document}
\title{Magnon-induced non-Markovian friction of a domain wall in a ferromagnet}

\author{Se Kwon Kim}
\affiliation{Department of Physics and Astronomy, University of California, Los Angeles, California 90095, USA}

\author{Oleg Tchernyshyov}
\affiliation{Department of Physics and Astronomy, Johns Hopkins University, Baltimore, Maryland 21218, USA}

\author{Victor Galitski}
\affiliation{Joint Quantum Institute and Condensed Matter Theory Center, Department of Physics, University of Maryland, College Park, Maryland 20742-4111, USA}

\author{Yaroslav Tserkovnyak}
\affiliation{Department of Physics and Astronomy, University of California, Los Angeles, California 90095, USA}

\begin{abstract}
Motivated by the recent study on the quasiparticle-induced friction of solitons in superfluids, we theoretically study magnon-induced intrinsic friction of a domain wall in a one-dimensional ferromagnet. To this end, we start by obtaining the hitherto overlooked dissipative interaction of a domain wall and its quantum magnon bath to linear order in the domain-wall velocity and to quadratic order in magnon fields. An exact expression for the pertinent scattering matrix is obtained with the aid of supersymmetric quantum mechanics. We then derive the magnon-induced frictional force on a domain wall in two different frameworks: time-dependent perturbation theory in quantum mechanics and the Keldysh formalism, which yield identical results. The latter, in particular, allows us to verify the fluctuation-dissipation theorem explicitly by providing both the frictional force and the correlator of the associated stochastic Langevin force. The potential for magnons induced by a domain wall is reflectionless, and thus the resultant frictional force is non-Markovian similarly to the case of solitons in superfluids. They share an intriguing connection to the Abraham-Lorentz force that is well-known for its causality paradox. The dynamical responses of a domain wall are studied under a few simple circumstances, where the non-Markovian nature of the frictional force can be probed experimentally. Our work, in conjunction with the previous study on solitons in superfluids, shows that the macroscopic frictional force on solitons can serve as an effective probe of the microscopic degrees of freedom of the system.
\end{abstract}

\date{\today}
\maketitle

\section{Introduction}

Solitons, stable nonlinear solutions in continuous fields theories, and their interactions with collective excitations have attracted much attention in a broad range of fields such as particle physics,~\cite{*[][{, and references therein.}] Manton} optics,~\cite{*[][{, and references therein.}] Taylor} and condensed matter physics~\cite{*[][{, and references therein.}] CL} because of fundamental interest as well as practical applications. Quasiparticles, elementary quanta of collective excitations, experience effective forces induced by the background solitons. The dynamics of solitons are, in return, influenced by the quasiparticles scattering with them. In particular, at finite temperatures, the thermal bath of quasiparticles can generate a deterministic frictional force and a stochastic Langevin force on a soliton. These two forces are caused by the same microscopic degrees of freedom, and, for that reason, are linked by the general relationship, which is manifested through the fluctuation-dissipation theorem.~\cite{CallenPR1951, KuboRPP1966}

Recently, \textcite{EfimkinPRL2016}, including one of us, have studied the frictional force experienced by a bright soliton in one-dimensional superfluids due to its interaction with Bogoliubov quasiparticles. The Ohmic friction that is linear in the velocity of the soliton is absent due to the integrability of the considered system. Instead, the frictional force is nonlocal in time and super-Ohmic in the low-frequency regime, about which the authors made an intriguing connection to the Abraham-Lorentz force that has been known in the classical electrodynamics for its causality paradox.~\cite{AbrahamAP1903, *LorentzPKNAW1904, Jackson} In addition, by using the Keldysh formalism,~\cite{KeldyshZETF1964, *KamenevAP2009} they obtained analytical expressions for the quasiparticle-induced frictional force and stochastic Langevin force on equal footing, which allows them to explicitly verify the fluctuation-dissipation theorem.

\begin{figure}
\includegraphics[width=\columnwidth]{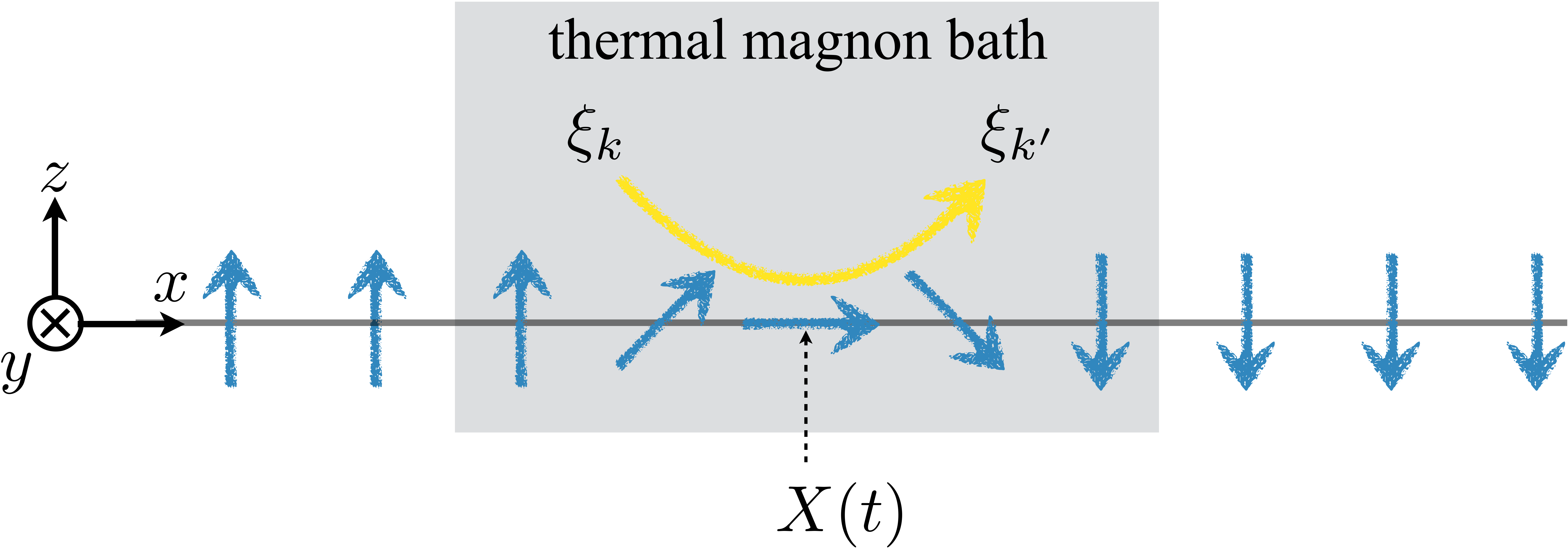}
\caption{Schematic illustration of a ferromagnetic domain wall and its thermal magnon bath, which induces a non-Markovian frictional force and a colored stochastic Langevin force on the domain wall. $X(t)$ represents the position of the domain wall; $\xi_k$ parametrizes the amplitude of magnon fluctuations. See the main text for details.}
\label{fig:fig1}
\end{figure}

Motivated by this study on solitons in superfluids, we investigate an analogous problem in magnetism: the magnon-induced frictional force on a domain wall in one-dimensional ferromagnets.~\cite{StampPRL1991, ChudnovskyPRB1992, BraunPRB1996} See Fig.~\ref{fig:fig1} for a schematic illustration. Specifically, we first identify a Berry-phase-induced coupling between a domain wall and its thermal magnon bath, with a focus on dissipative effects that have been overlooked heretofore. Then, by integrating out the magnon bath, we derive the following generalized Langevin equation~\cite{KuboRPP1966, HanggiRMP1990, *HanggiChaos2005, Peters} for the dynamics of a domain wall:
\begin{equation}
M \ddot{X}(t) + \int_{-\infty}^t dt' \eta(t - t') \dot{X} (t') = F(t) + \zeta(t) \, ,
\label{eq:langevin}
\end{equation}
where the frictional force, the second term on the left-hand side, is induced by the aforementioned coupling. Here, $X(t)$ and $M$ are the position and the effective mass of a domain wall; $\eta(t)$ is the retarded response kernel;~\footnote{Since the response kernel $\eta(t)$ is used only with positive argument $t > 0$ in the generalized Langevin equation, we can assume, without loss of generality, that it is even in time, $\eta(t) = \eta(-t)$, or that it is positive $\eta(t) > 0$ only for $t > 0$. Although we use the latter, one can use the former, e.g., as in \textcite{EfimkinPRL2016}.} $F(t)$ and $\zeta(t)$ are the external and the stochastic (Langevin) forces on a domain wall. The deterministic response kernel and the stochastic force are linked by the quantum fluctuation-dissipation theorem via the spectral function $J(\omega)$ as follows:~\cite{CallenPR1951, KuboRPP1966, HanggiRMP1990}
\begin{eqnarray}
\eta(t) &=& \frac{2 \Theta(t)}{\pi} \int_0^\infty d\omega \, J(\omega) \cos (\omega t) \, , \label{eq:eta} \\
\langle \zeta(t) \zeta(0) \rangle &=& \frac{1}{\pi} \int_0^\infty d\omega \, \frac{ \hbar \omega J(\omega)}{\tanh (\hbar \omega / 2 T)} \cos (\omega t) \, , \label{eq:zeta}
\end{eqnarray}
where $\Theta(t)$ is the Heavyside step function. Here, the spectral function $J(\omega)$ is the real dissipative part of the Fourier transform of the response kernel, $J(\omega) =\text{Re} \, \eta[\omega]$.~\footnote{The real part of $\eta[\omega]$ is even in frequency and thus describes dissipative effects. Its imaginary part $\text{Im} \, \eta[\omega]$ is, on the other hand, odd in frequency and therefore give rise to non-dissipative effects such as mass renormalization.} In the classical limit, where the temperature is much higher than the characteristic frequency of the dynamics, $T \gg \hbar \omega$, they satisfy the classical fluctuation-dissipation theorem: $\langle \zeta(t) \zeta(0) \rangle = T \eta \left( |t| \right)$.~\cite{HanggiRMP1990, FariasPRE2009, *RuckriegelPRL2015} 

There is no reflection of magnons scattering off a domain wall,~\cite{WinterPR1961, *ThielePRB1973, *WangPRB2014} and thus there is no Ohmic frictional force.~\cite{FedichevPRA1999, EfimkinPRL2016} However, a finite viscous force is induced by backreaction of thermal magnons that are perturbed by the domain-wall acceleration (i.e., temporal variation of the domain-wall velocity), analogous to the reactive effects of electromagnetic radiation on the motion of a charged particle.~\cite{Jackson} The resultant viscous force is non-Markovian and super-Ohmic in the low-frequency regime similarly to the case of superfluid solitons.~\footnote{We call a friction Markovian subjected to the approximation $\gamma \dot{X} \approx \int^t_{-\infty} dt' \eta(t - t') \dot{X}(t')$,~\cite{Peters} i.e., the response kernel can be approximated by the delta function, $\eta(t) \approx 2 \gamma \delta(t)$. The magnon-induced friction obtained in our work is not Markovian since it vanishes in the zero-frequency limit.} Although the predicted phenomenon is similar, our theory is simplified in comparison to the superfluid counterpart in the appropriate limit, where the quasiparticle-soliton scattering process preserves the number of quasiparticles.

The paper is organized as follows. In Sec.~\ref{sec:model}, we obtain exact solutions for a domain wall with spin waves on top of it in a one-dimensional ferromagnet. We then derive their interaction to linear order in the domain-wall velocity and quadratic order in spin-wave amplitudes, which stems from the spin Berry-phase term in the Lagrangian. The identification of this interaction, which can be found in Eq.~(\ref{eq:L-coupling}), is our first main result. In Sec.~\ref{sec:friction}, by treating a domain wall as a classical particle embedded in a magnonic quantum bath, we derive the expression for the magnon-induced response kernel $\eta(t)$, which is our second main result that can be found in Eq.~(\ref{eq:eta-2}), in two different ways. First, we employ the time-dependent perturbation theory in quantum mechanics~\cite{LL3} to obtain the frictional force. This approach helps us understand the force's origin in that a domain wall loses its energy to thermal magnons via inelastic scattering. Second, we derive the Langevin equation within the Keldysh formalism,~\cite{KeldyshZETF1964} which allows us to obtain explicit expressions for both the frictional force and the stochastic Langevin force on equal footing and thereby to verify the fluctuation-dissipation theorem. Two independent approaches yield identical results. In Sec.~\ref{sec:phenomena}, we discuss dynamic responses to an oscillating force and experimental prospects to probe it. In Sec.~\ref{sec:discussion}, we compare our results with those of the existing literature~\cite{StampPRL1991, ChudnovskyPRB1992, BraunPRB1996} and discuss key approximations made in our model. In Sec.~\ref{sec:outlook}, we conclude the paper by discussing an outlook on future work.

\section{A domain wall and spin waves}
\label{sec:model}

Our model system is a one-dimensional ferromagnet, which is described by the following Hamiltonian:
\begin{equation}
H = \int dx \, \left[ A \mathbf{n}'^2 + K_e (1 - n_z^2) + K_h n_y^2 \right] / 2 \, ,
\label{eq:H}
\end{equation}
where the three-dimensional unit vector $\mathbf{n}$ represents the direction of the local magnetization and $'$ is the spatial gradient in the $x$ direction. Here, the positive coefficients $A$ and $K_e$ parametrize the exchange stiffness and easy-axis anisotropy along the $z$ axis, respectively, and the nonnegative coefficient $K_h$ parametrizes hard-axis anisotropy along the $y$ axis. The dynamics of the magnet can be described by the following Lagrangian:
\begin{equation}
L = - s \int dx \, \mathbf{a} (\mathbf{n}) \cdot \dot{\mathbf{n}} - H \, ,
\label{eq:L}
\end{equation}
where $s$ is the spin density per unit length and $\mathbf{a}$ is the vector potential for a magnetic monopole, $\boldsymbol{\nabla}_\mathbf{n} \times \mathbf{a} (\mathbf{n}) = \mathbf{n}$. The first term accounts for the effects of the spin Berry phase, which governs the dynamics of the magnet.~\cite{Altland}

The ferromagnet has two ground states, $\mathbf{n} \equiv \pm \hat{\mathbf{z}}$, which are uniformly polarized along the easy axis. A domain wall is a solution that minimizes the Hamiltonian $H$ for boundary conditions, $\mathbf{n} (x = \pm \infty) = \pm \hat{\mathbf{z}}$, which is given by
\begin{equation}
\cos \theta_0 = \tanh \left( \frac{x - X}{\lambda} \right) \, , \quad \phi_0 \equiv \Phi \, ,
\end{equation}
where $\lambda \equiv \sqrt{A/K}$ is the width of the domain wall, $\theta$ and $\phi$ are the polar and the azimuthal angles in the spherical representation of $\mathbf{n} = (\sin \theta \cos \phi, \sin \theta \sin \phi, \cos \theta)$.~\cite{SchryerJAP1974} Here, $X$ is the position of the domain wall, which parametrizes the zero-energy mode associated with the spontaneous breaking of the continuous translational symmetry; $\Phi$ is the azimuthal angle of the domain wall, which is either $0$ or $\pi$ in the presence of the hard-axis anisotropy, $K_h > 0$. In the absence of the anisotropy, $K_h = 0$, the Hamiltonian $H$ is invariant under spin rotations about the $z$ axis, and $\Phi$ becomes the parameter for the zero-energy mode associated with the spontaneous breaking of this continuous spin-rotational symmetry.

To simplify the subsequent discussions, we use natural units of length, time, and energy, which are given by
\begin{equation}
\lambda = \sqrt{A/K_e} \, , \quad \tau \equiv s / K_e \, , \quad \epsilon \equiv \sqrt{A K_e} \, ,
\label{eq:unit}
\end{equation}
respectively. These parameters have natural interpretations in terms of domain-wall characteristics: $\lambda$ is the domain-wall width, $2 \epsilon$ is the domain-wall rest energy; $c \equiv \lambda / \tau$ is the domain-wall velocity for Cherenkov radiation.~\cite{BaryakhtarSPU1985, *YanAPL2011} Also note that the product of the energy and the time scales is given by $\epsilon \cdot \tau = s \lambda$, which represents the total spin contained within the domain-wall width. The time scale $\tau$ also sets the energy scale of a magnon as will be shown below. Using these scales amounts to setting the parameters, $A, K_e,$ and $s$ to $1$ and replacing the hard-axis anisotropy coefficient $K_h$ by a dimensionless number $\kappa \equiv K_h / K_e$. In this paper, unless specified, we do not consider the extrinsic damping of spin dynamics that can arise due to the coupling to the nonmagnetic degrees of freedom, such as phonons or electrons.

\subsection{Spin waves on a static domain wall}

The exact solutions of spin-wave modes on a static domain wall are known,~\cite{WinterPR1961, KimPRB2014} which we present below. To simplify calculations, we set $X = 0$ and $\Phi = 0$ in this section without loss of generality. We start by expanding the Lagrangian to the quadratic order in the deviations from the domain-wall solution, which we describe by two variables: $\delta n_1 = \delta \theta$ represents the change of the magnetization in the plane of the domain-wall spin texture and $\delta n_2 = \sin \theta_0 \delta \phi$ represents the change out of the plane. The first-order term is absent because the domain wall is a stationary solution to the equations of motion. The second-order term is given by
\begin{equation}
L_2 = - \int dx ( \delta n_1 \delta \dot{n}_2 + \delta n_1 \mathcal{H}_1 \delta n_1 / 2 + \delta n_2 \mathcal{H}_2 \delta n_2 / 2 ) \, ,
\label{eq:L2}
\end{equation}
where the first term is from the spin Berry phase and the second and third terms are from the Hamiltonian. Here, the Hamiltonian densities are given by
\begin{eqnarray}
\mathcal{H}_1 &=& - \frac{d^2}{dx^2} + [ 1 - 2 \sech^2 (x) ] = a^\dagger a \, , \\
\mathcal{H}_2 &=& - \frac{d^2}{dx^2} + [ 1 + \kappa - 2 \sech^2 (x) ] = a^\dagger a + \kappa \, , \label{eq:H2}
\end{eqnarray}
where $a \equiv d / dx + \tanh x$, and $a^\dagger \equiv - d / dx + \tanh x$. The equation of motion for spin waves is given by the following ``Schr{\"o}dinger equation":
\begin{equation}
\label{eq:se}
\frac{d}{dt} \begin{pmatrix} \delta n_1 \\ \delta n_2 \end{pmatrix} 
= 
\begin{pmatrix} 0 & \mathcal{H}_1 \\ - \mathcal{H}_2 & 0 \end{pmatrix} \begin{pmatrix} \delta n_1 \\ \delta n_2 \end{pmatrix} \, .
\end{equation}
Note that the Hamiltonian densities include the spatially varying potentials as a consequence of the translational symmetry breaking due to a domain wall. The potential $U (x) = - 2 \sech^2 (x)$, which is named after \textcite{PoschlZP1933}, has a remarkable property: waves pass through it without any reflection as shown below. 

With the aid of the technique of supersymmetric quantum mechanics,~\cite{CooperPR1995} the above Hamiltonian densities can be related to the following simpler ones:
\begin{eqnarray}
\mathcal{H}^0_1 &=& - \frac{d^2}{dx^2} + 1 = a a^\dagger \, , \\
\mathcal{H}^0_2 &=& - \frac{d^2}{dx^2} + 1 + \kappa = a a^\dagger + \kappa \, .
\end{eqnarray}
These Hamiltonian densities are translationally invariant and thus can be diagonalized by expanding the fields in terms of plane waves $\propto \exp(i k x)$. The corresponding spin-wave modes are those of a uniform ground state and they are elliptical in the presence of the hard-axis anisotropy, $\kappa > 0$. Continuum solutions to the original problem can be obtained by applying the operator $a^\dagger$ to these plane-wave solutions and it can be shown that they share the same frequency. 

The resultant continuum solutions to Eq.~(\ref{eq:se}) can be summarized as follows:
\begin{eqnarray}
\delta n_1 (x, t) &=& \int \frac{d k}{2 \pi} \sqrt{2 \hbar} c_k \text{Re} \left[ \xi_k (t) \psi_k (x) e^{-i \omega_k t} \right] \, , \label{eq:n1} \\
\delta n_2 (x, t) &=& \int \frac{d k}{2 \pi} \sqrt{2 \hbar} c_k^{-1} \text{Im} \left[ \xi_k (t) \psi_k (x) e^{-i \omega_k t} \right] \, , \label{eq:n2}
\end{eqnarray}
where $c_k = [(1 + k^2 + \kappa) / (1 + k^2)]^{1/4}$ represents the ellipticity of a spin wave mode at momentum $k$, 
\begin{equation}
\label{eq:dispersion}
\omega_k = \sqrt{(1 + k^2)(1 + k^2 + \kappa)}
\end{equation}
is the frequency at momentum $k$, and $\xi_k(t)$ is the complex-valued amplitude that varies slowly on the time scale set by the spin-wave gap $\omega_0$. The complex-valued function $\psi_k (x)$ is given by
\begin{equation}
\label{eq:psi-k}
\psi_k (x) = \frac{a^\dagger}{1 - ik} e^{i k x} = \frac{\tanh(x) - i k}{1 - i k} e^{i k x} \, ,
\end{equation}
which satisfies the orthogonality condition, $\int dx \psi_k^* (x) \psi_{k'} (x) = 2 \pi \delta (k - k')$. With the above solutions, the second-order Lagrangian term $L_2$ is given by
\begin{equation}
L_2 = \int \frac{d k}{2 \pi} \left( i \hbar \xi_k^* \dot{\xi}_k - \epsilon_k \xi_k^* \xi_k \right) \, ,
\end{equation}
where $\epsilon_k \equiv \hbar \omega_k$ is the magnon energy at momentum $k$.

Besides the continuum modes, there are also two localized modes, $\delta n_1 \propto \partial_x \theta_0 (x) = - \sech (x)$ and $\delta n_2 \propto \sin \theta_0 (x) = \sech (x)$, which correspond to the change of the magnetization upon infinitesimal domain-wall displacement, $X \mapsto X + \delta X$, and rotation, $\Phi \mapsto \Phi + \delta \Phi$, respectively. The dynamics of these modes will be treated explicitly in the next section by promoting $X$ and $\Phi$ to dynamic variables. 

\subsection{Spin waves on a moving domain wall}

In order to study the interaction between the domain-wall motion and spin waves, we allow the domain-wall position and angle variables to be time dependent, $X(t)$ and $\Phi(t)$. Since our primary interest is on the coupling of the translational motion of the domain wall and spin waves, we henceforth focus on the case of a finite hard-axis anisotropy, $\kappa > 0$, in which the angle variable $\Phi$ becomes a slave mode of the position variable $X$ as will be shown below. By using the expressions $\theta(x, t) = \theta_0 \left( x - X(t) \right) + \delta \theta \left( x - X(t), t \right)$ and $\phi (x, t) = \Phi(t) + \delta \phi \left( x - X(t), t \right)$ in the Lagrangian $L$ [Eq.~(\ref{eq:L})] and expanding it to the linear order in $\dot{X}$ and the quadratic order in $\delta n_1, \delta n_2$, and $\Phi$, we obtain the following terms:
\begin{equation}
\label{eq:L-coupling}
\begin{split}
L = & 2 \Phi \dot{X} - \kappa \Phi^2 \\
& + \int \frac{dk}{2 \pi} \left[ i \hbar \xi_k^* \dot{\xi}_k - \epsilon_k \xi_k^* \xi_k \right] \\
&+ \dot{X} P_m \, ,
\end{split}
\end{equation}
where $P_m$ is the total momentum of spin waves given by~\footnote{The natural unit of linear momentum is given by $\epsilon / c = s$, the angular momentum density per unit length.}
\begin{equation}
\label{eq:Pm}
P_m = \int \frac{dk dk'}{(2 \pi)^2} \pi_{k k'} \xi_k^* \xi_{k'} e^{i (\omega_k - \omega_{k'}) t} \, , 
\end{equation}
with the momentum-space hopping amplitude
\begin{equation}
\label{eq:pi}
\pi_{k k'} = \frac{1}{2} \left( \frac{c_k}{c_{k'}} + \frac{c_{k'}}{c_k} \right) \int dx \, \psi_k^* (- i \hbar \partial_x) \psi_{k'} \, .
\end{equation}
Here, we expanded $\delta \theta$ and $\delta \phi$ as the linear combinations of the continuum spin-wave modes only by excluding the zero modes, and disregarded the rapidly oscillating terms at frequencies higher than the spin-wave gap $\omega_0$ by focusing on slow dynamics. There is no linear term in the spin-wave field $\xi_k$ due to the orthogonality of the spin-wave eigenstates. 

In Eq.~(\ref{eq:L-coupling}), the right-hand side on the first line describes the dynamics of the two generalized coordinates of the domain wall, $X$ and $\Phi$,~\footnote{We obtain the Lagrangian term for the zero-mode dynamics by using the following expression for the spin Berry phase: $- s \mathbf{a}(\mathbf{n}) \cdot \dot{\mathbf{n}} = - s (1 - \cos \theta) \dot{\phi}$, in which a gauge is chosen for a magnetic-monopole vector potential. The resultant term is $- 2 X \dot{\Phi}$, which we transform to $2 \Phi \dot{X}$ by performing the integration by parts over time $t$ and dropping the total-derivative term.} from which we can obtain the equation of motion for $\Phi$: $\Phi = \dot{X} / \kappa$.~\cite{SchryerJAP1974} As stated earlier, $\Phi (t)$ is completely determined by $X(t)$ and thus is a slave mode of it. Replacing $\Phi$ by $\dot{X} / \kappa$ transforms the right-hand side on the first line to the domain-wall kinetic energy $\dot{X}^2 / \kappa$, which leads us to identify $2 / \kappa$ as the effective mass $M$ of the domain wall. The second line describes continuum spin-wave modes. The third line represents the linear coupling between the velocity of the domain wall and the total momentum of spin waves. This coupling is our first main result and constitutes the important starting point for the subsequent development of the theory of quantum friction of the domain wall. The total momentum of the system is given by 
\begin{equation}
P = \frac{d L}{d \dot{X}} = M \dot{X} + P_m \, ,
\label{eq:P}
\end{equation}
where the first and second terms are the contributions from the domain wall and spin waves, respectively. Both contributions are rooted in the spin Berry-phase term in the Lagrangian, which has already been identified as a source of the linear momentum of a ferromagnet.~\cite{HaldanePRL1986, *VolovikJPC1987, *YanPRB2013, *WongPRB2009, TchernyshyovAP2015} In Appendix~\ref{app:conserved}, we discuss an alternative way to derive the total momentum $P$ using the collective-coordinate approach.~\cite{TretiakovPRL2008}

By using the spin-wave solutions [Eq.~(\ref{eq:psi-k})], we can obtain the exact expression for the momentum-space hopping amplitude $\pi_{k k'}$ [Eq.~(\ref{eq:pi})] whose last integral factor is given by
\begin{equation}
\begin{split}
\int dx \, \psi_k^*& (- i \hbar \partial_x) \psi_{k'} = 2 \pi \hbar k \, \delta (k - k') \\
& + \frac{\pi \hbar}{2 \sinh (\pi (k - k') / 2)} \frac{k^2 - k'^2}{(1 - i k') (1 + ik)} \, .
\end{split}
\end{equation}
Let us make a few remarks on the hopping amplitude $\pi_{k k'}$. First, it is Hermitian, $\pi_{k k'} = \pi^*_{k' k}$. Second, it has off-diagonal components, reflecting the breaking of the translational symmetry due to the domain wall. Third, the backscattering is absent, $\pi_{k, -k} = 0$, similarly to the case of Bogoliubov quasiparticles on top of superfluid solitons,~\cite{EfimkinPRL2016} which stems from the integrability of both systems.

\section{Magnon-induced friction}
\label{sec:friction}

In this section, we derive the frictional force on the domain wall, which is induced by its coupling to thermally excited spin waves. The domain wall is treated as a heavy semiclassical object, whereas thermal magnons, quanta of spin waves, are considered to be light and form a thermal bath for the domain wall, the justification of which is given in Appendix~\ref{app:mass}. To this end, we take two different approaches: time-dependent perturbation theory in quantum mechanics and Keldysh formalism. Two approaches are complementary. The former allows us to obtain the frictional force with a clear physical picture of its origin, but it does not provide the fluctuation properties of the stochastic force directly. The latter approach is computationally more demanding and thus it can be more difficult to understand the physical origin of the friction within it. However, the Keldysh technique is powerful: it yields the frictional force and the correlator of the stochastic force on equal footing, which allows us to check for internal consistency. We will see below that the magnon-induced friction is induced by the backreaction of the thermal magnon gas perturbed by the domain-wall acceleration, which is analogous to the reactive effects of electromagnetic radiation on the equations of motion of a charged particle.~\cite{Jackson} We here focus on two-magnon scattering with a domain wall while neglecting other scattering processes involving more magnons by assuming magnons are sufficiently dilute.

\subsection{Time-dependent perturbation theory}

Let us first derive the frictional force by using the time-dependent perturbation theory. Specifically, we seek the expression for the response kernel $\eta(t)$ in the Langevin equation~(\ref{eq:langevin}). To treat thermal magnons as the quantum bath, we quantize spin waves by promoting the complex scalar fields $( \xi_k \, , \xi_k^* )$, which is the pair of canonically conjugate variables, to the magnon annihilation and creation operators, $( \hat{\xi}_k \, , \hat{\xi}_k^\dagger )$. The term $V(t) P_m$ in the Lagrangian [Eq.~(\ref{eq:L-coupling})] that couples the domain wall velocity $V(t)$ and the magnon bath can be interpreted as a time-dependent term in the Hamiltonian,
\begin{equation}
\hat{W}(t) = - V(t) \int \frac{dk dk'}{(2 \pi)^2} \pi_{k k'} \hat{\xi}_k^\dagger \hat{\xi}_{k'} e^{i (\epsilon_{k} - \epsilon_{k'}) t / \hbar} \, .
\label{eq:W}
\end{equation}
By treating this term as a domain-wall-induced perturbation on the magnonic Hamiltonian within the time-dependent perturbation theory,~\cite{LL3} we can derive the probability that a magnon at the state $k$ is found to be at the state $k'$ after time $t$, which, in its leading order, is given by
\begin{equation}
P_{k k'} (t) = \frac{1}{\hbar^2} \left| \int_0^t dt' e^{i (\epsilon_k - \epsilon_{k'}) t' / \hbar} \pi_{k k'} V(t') \right|^2 \, .
\end{equation}
The transition rate of one magnon from $k$ to $k'$ is the time derivative of $P_{k k'} (t)$, which is given by
\begin{equation}
\label{eq:R}
\begin{split}
R_{k k'} (t) =& \frac{2 |\pi_{k k'}|^2 V(t)}{\hbar^2} \\
&\times \int_0^t dt' \, \cos\left[ \frac{(\epsilon_{k'} - \epsilon_k) (t - t')}{\hbar} \right] V(t') \, .
\end{split}
\end{equation}
The transition rate is symmetric with respect to the momentum exchange, $R_{k k'} (t) = R_{k' k} (t)$. Note that the foregoing history of the domain-wall motion influences the transition rate of a magnon via the last integral factor.

To derive the frictional force on the domain wall, we now consider the energy gain of magnons during their scattering off the domain wall, which is, by the conservation of the total energy, identical to the energy loss of the domain wall. Using the previous result on the magnon transition rate, the energy-dissipation rate from the domain wall to the magnon bath is given by
\begin{eqnarray}
P &=& \int \frac{dk dk'}{(2 \pi)^2} R_{k k'} (t) (\epsilon_{k'} - \epsilon_k) f_k \, , \label{eq:P1} \\
 &=& \frac{1}{2} \int \frac{dk dk'}{(2 \pi)^2} R_{k k'} (t) (\epsilon_{k'} - \epsilon_k) (f_k - f_{k'}) \, , 
\end{eqnarray}
where $f_k \equiv 1/[\exp(\epsilon_k / T) - 1]$ is the Bose-Einstein distribution function at momentum $k$. On the other hand, from the Langevin equation~(\ref{eq:langevin}) for the domain wall, the energy dissipation rate is given by
\begin{equation}
P = V(t) \int_0^t dt' \eta(t - t') V(t') \, .
\label{eq:P2}
\end{equation}
By matching Eq.~(\ref{eq:P2}) to Eq.~(\ref{eq:P1}) in conjunction with Eq.~(\ref{eq:R}), we can obtain the expression for the response kernel:
\begin{equation}
\label{eq:eta-2}
\begin{split}
\eta \left( \Delta t \right) =\frac{\Theta(t)}{\hbar^2} \int \frac{dk dk'}{(2 \pi)^2} & |\pi_{k k'}|^2 (\epsilon_{k'} - \epsilon_k) (f_k - f_{k'}) \\
& \times \cos\left[ \frac{(\epsilon_{k'} - \epsilon_k) \Delta t}{\hbar} \right] \, .
\end{split}
\end{equation}
The corresponding spectral function $J(\omega)$ in Eq.~(\ref{eq:eta}) can be obtained by the Fourier transformation: 
\begin{equation}
J(\omega) = \frac{\pi \omega}{2} \int \frac{dk dk'}{(2 \pi)^2} |\pi_{k k'}|^2 (f_{k} - f_{k'}) \delta [\hbar \omega - (\epsilon_{k'} - \epsilon_k)] \, ,
\label{eq:J}
\end{equation}
which can be considered as a manifestation of Fermi's golden rule.~\cite{LL3} The time-dependent perturbation theory allows us to obtain the response kernel, but not the autocorrelation of the stochastic force $\zeta$. We, however, can invoke the fluctuation-dissipation theorem to obtain it.~\cite{KuboRPP1966} The physical picture of the emergence of the frictional force is the following: Via the interaction term $W$ [Eq.~(\ref{eq:W})], magnons absorb the part of the domain-wall energy, which, in return, gives rise to the frictional force on the domain wall. Instead of the conservation of the total energy invoked above, we can alternatively use the conservation of the total linear momentum to obtain the same result, the details of which can be found in Appendix~\ref{app:momentum}.

Figure~\ref{fig:fig2} shows the plots of the response kernel $\eta(t)$ [Eq.~(\ref{eq:eta-2})] and the real part of its Fourier transform $J(\omega) = \text{Re} \, \eta[\omega]$ at the temperature $T = \epsilon_0 / 10$ and in the vanishing hard-axis anisotropy limit $\kappa = 0$. The dissipation is nonlocal in time and thus the associated stochastic force should have a colored noise. In the limit of low frequency and low temperature, $\hbar \omega \ll T \ll \epsilon_0$, the response kernel can be approximated by a super-Ohmic one as follows:
\begin{equation}
\label{eq:eta-approx}
J(\omega) \simeq \frac{\hbar (1 + \kappa) e^{- \epsilon_0 / T}}{\pi (2 + \kappa)^2 \lambda^2} (\omega \tau)^2 \equiv \eta_0 (\omega \tau)^2 \, ,
\end{equation}
which is written in physical units instead of natural units for transparent interpretation. Since the response kernel is second order in time derivative, the corresponding frictional force is third order in time derivative: $F_\text{AL} \propto \dddot{X}$. This force $F_\text{AL}$ is known as the Abraham-Lorentz force, which has been studied in the classical electrodynamics of a charged particle coupled with its own radiation.~\cite{AbrahamAP1903} The Abraham-Lorentz force is famous for causing the causality paradox, invalidating the approximation taken above. See Sec.~\ref{sec:well} for more discussions on the Abraham-Lorentz force. In addition, note that the response kernel $J(\omega)$ vanishes in the classical limit $\hbar \rightarrow 0$, which indicates the quantum nature of the origin of the corresponding frictional force.

\begin{figure}
\includegraphics[width=\columnwidth]{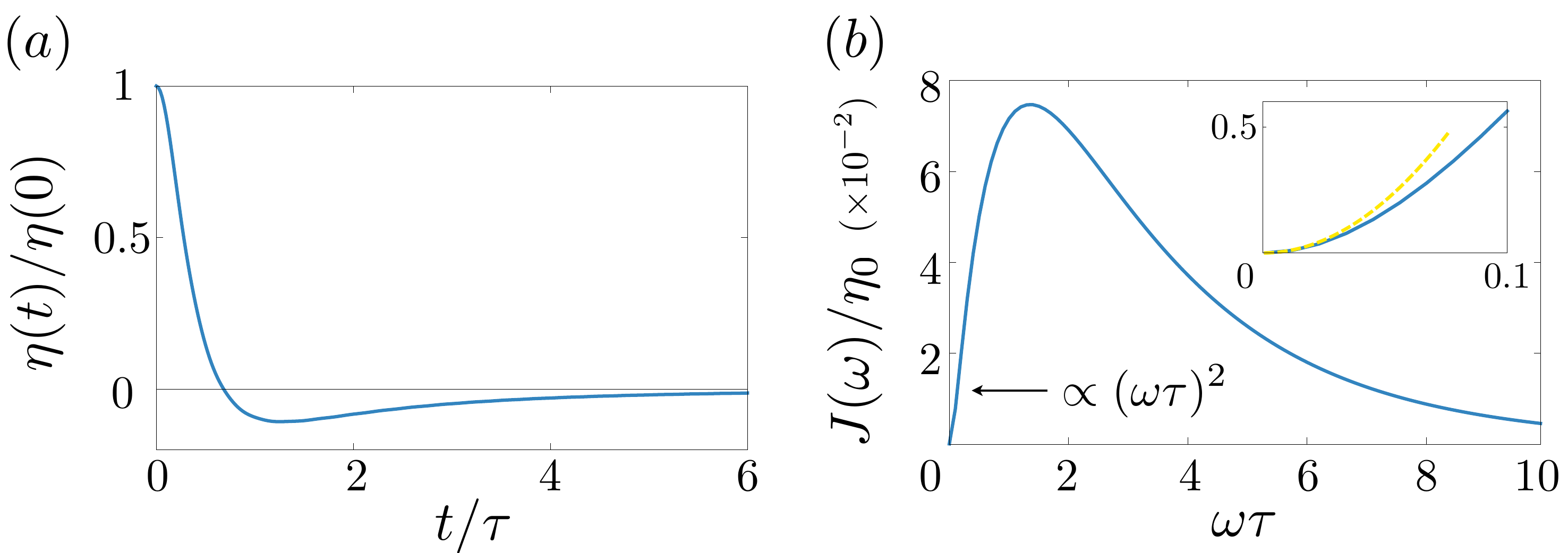}
\caption{Plots of a response kernel in (a) the time domain, $\eta(t)$, and (b) the frequency domain, $J(\omega) = \text{Re} \, \eta[\omega]$ at temperature $T = \epsilon_0 / 10$. The solid blue lines show the numerical evaluations of Eqs.~(\ref{eq:eta-2}) and (\ref{eq:J}). The dashed yellow lines in the inset show the analytical expression for the low-frequency limit of $J(\omega)$ in Eq.~(\ref{eq:eta-approx}).}
\label{fig:fig2}
\end{figure}

\subsection{Keldysh formalism}

In this section, we use the Keldysh technique~\cite{KeldyshZETF1964} to derive both the frictional force and the stochastic Langevin force. Let us first write down the action:
\begin{equation}
\begin{split}
S = & \frac{M}{2} \int dt \dot{X}^2 + \int dt \int \frac{dk}{2 \pi} \left[ i \hbar \xi_k^* \dot{\xi}_k - \epsilon_k \xi_k^* \xi_k \right] \\
& + \dot{X} \int dt \int \frac{dk dk'}{(2 \pi)^2} \pi_{k k'} \xi_k^* \xi_{k'} e^{i (\epsilon_{k} - \epsilon_{k'}) t / \hbar} \, ,
\end{split}
\end{equation}
which is obtained from the Lagrangian $L$ [Eq.~(\ref{eq:L-coupling})] after replacing $\Phi$ with its classical solution $\dot{X} / \kappa = \dot{X} / 2 M$. By closely following the approach taken by \textcite{EfimkinPRL2016} for solitons in superfluids, we will obtain below the quasiclassical equations of motion for $X$ as the saddle point of a one-loop effective action in the Keldysh formalism, which corresponds to an expansion of the action to the quadratic order in $\dot{X} \ll 1$.

We start by duplicating the dynamic degrees of freedom, $X \rightarrow X_+ \, , X_-$ and $\xi_k \rightarrow \xi_{+, k} \, , \xi_{-, k}$, where the variables with subscripts $+$ and $-$ reside on the forward and the backward parts of the Keldysh contour, respectively. The classical degrees of freedom are given by the average of the duplicated fields: $X_c \equiv (X_+ + X_-) / 2$ and $\xi_{c, k} \equiv (\xi_{+, k} + \xi_{-, k}) / \sqrt{2}$. The quantum degrees of freedom are given by the difference between them: $X_q \equiv (X_+ - X_-)/2$ and $\xi_{q, k} \equiv (\xi_{+, k} - \xi_{-, k}) / \sqrt{2}$.~\footnote{For the spin-wave field transformations, we followed the convention used by \textcite{KamenevAP2009}, which involves $1/\sqrt{2}$ rather than $1/2$. For the position-variable transformations, we used $X_q \equiv (X_+ - X_-) / 2$ instead of $X_q \equiv X_+ - X_-$ to simplify the expression for the Keldysh action in Eq.~(\ref{eq:SK}).}
In terms of the classical and the quantum components, the Keldysh action is given by
\begin{equation}
\label{eq:SK}
\begin{split}
S^K = & 2 M \int dt \dot{X}_c \dot{X}_q \\
+ & \int dt \int dt' \int \frac{dk}{2 \pi} \boldsymbol{\xi}_k^\dagger (t) \hat{G}^{-1}_k (t, t') \boldsymbol{\xi}_k (t') \\
+ & \int dt \int \frac{dk dk'}{(2 \pi)^2} \pi_{k k'} \boldsymbol{\xi}_k^\dagger (t) \hat{\dot{X}} \boldsymbol{\xi}_{k'} (t) e^{i (\epsilon_{k} - \epsilon_{k'}) t / \hbar} \, ,
\end{split}
\end{equation}
where $\boldsymbol{\xi}_k \equiv (\xi_{c,k}, \xi_{q, k})$, $\dagger$ stands for the Hermitian conjugation,
\begin{equation}
\begin{split}
\hat{G}_k (t, t') &\equiv -i \langle \boldsymbol{\xi}_k (t) \boldsymbol{\xi}_k^\dagger (t') \rangle \\
& \equiv \begin{pmatrix} G^K_k (t, t') & G^R_k (t, t') \\ G^A_k (t, t') & 0 \end{pmatrix} \\
& = -i e^{- i \epsilon_k (t - t') / \hbar} \begin{pmatrix} (1 + 2 f_k ) & \Theta(t - t') \\  - \Theta(t' - t) & 0 \end{pmatrix} \, ,
\end{split}
\label{eq:G}
\end{equation}
and
\begin{equation}
\hat{\dot{X}} = \begin{pmatrix} \dot{X}_q & \dot{X}_c \\ \dot{X}_c & \dot{X}_q \end{pmatrix} \, .
\end{equation}
Here, $G^R, G^A$, and $G^K$ are the retarded, advanced, and Keldysh Green functions, respectively. In the frequency domain, we have
\begin{equation}
\begin{split}
\hat{G}_k (\epsilon) & = \begin{pmatrix} G^K_k (\epsilon) & G^R_k (\epsilon) \\ G^A_k (\epsilon) & 0 \end{pmatrix} \\
& = \begin{pmatrix} - 2 \pi i (1 + 2 f_k) \delta(\epsilon - \epsilon_k) & (\epsilon - \epsilon_k + i 0^+)^{-1} \\ (\epsilon - \epsilon_k - i 0^+)^{-1} & 0 \end{pmatrix} \, ,
\end{split}
\end{equation}
where $0^+$ represents an infinitesimally small positive number. By using $f_k = 1/[\exp(\epsilon_k / T) - 1]$, we can explicitly check that the Keldysh Green functions satisfy $G^K (\epsilon) = [G^R (\epsilon) - G^A (\epsilon)] \coth (\epsilon / 2 T)$, which constitutes the statement of the fluctuation-dissipation theorem within the Keldysh formalism.

After integrating out the magnon modes, the details of which is in Appendix~\ref{app:Keldysh}, we obtain the following effective action for $X$:
\begin{equation}
\begin{split}
S^K_\text{eff} & = \int dt \left\{ 2 X_q (t) \left[ - M \ddot{X}_c  - \int_0^t dt' \eta(t - t') \dot{X}_c (t') + \zeta(t) \right] \right\} \\
& + \frac{i \hbar}{2} \int dt \int dt' \zeta (t) C_s^{-1} (t - t') \zeta (t') \, ,
\end{split}
\end{equation}
where $\eta(t)$ and $C_s (t - t') = \langle \zeta(t) \zeta(t') \rangle$ are given by Eqs.~(\ref{eq:eta}) and (\ref{eq:zeta}), respectively, with $J(\omega)$ in Eq.~(\ref{eq:J}). The results within the Keldysh formalism is identical to the previous ones obtained within the time-dependent perturbation theory. Here, we would like to comment on the origin of the stochastic force $\zeta(t)$. It is an auxiliary field introduced by the Hubbard-Stratonovich transformation~\cite{HubbardPRL1959} that is employed to remove the quadratic-order term in the quantum component $X_q$. In this sense, the stochastic Langevin force $\zeta(t)$ is rooted in the intrinsic fluctuations of the system. The saddle-point solution of the effective action with respect to the quantum variable $X_q$ gives the quasiclassical equation of motion for the coordinate $X_c$:
\begin{equation}
M \ddot{X}_c (t) + \int_{-\infty}^t dt' \eta(t - t') \dot{X}_c(t') = \zeta(t) \, ,
\end{equation}
which is identical to Eq.~(\ref{eq:langevin}) in the absence of an external force.

\section{Non-Markovian Friction}
\label{sec:phenomena}

We now discuss effects of the non-Markovian frictional force on the dynamical response of the domain wall. In this section, we include the local-in-time extrinsic frictional force and the associated white-noise stochastic force as additional terms in the equations of motion, which we will contrast with the magnon-induced non-Markovian friction. In addition, we now return to the physical units instead of the natural units [Eq.~(\ref{eq:unit})]. The resultant equation of motion for $X$ is given by 
\begin{equation}
M \ddot{X} + \int_0^t dt' \eta(t - t') \dot{X}(t') + \gamma \dot{X} = F(t) + \zeta(t) + \nu(t) \, .
\label{eq:eom}
\end{equation}
Here, $F(t)$ is an external force on the domain wall, $\gamma = 2 \alpha s / \lambda$ parametrizes the Markovian frictional force rooted in the local-in-time Gilbert damping,~\cite{GilbertIEEE2004} which is usually attributed to dissipation to phonon bath, and $\nu(t)$ is the associated stochastic Langevin force.~\cite{BrownPR1963, *KuboPTPS1970, IvanovJPCM1993} The Markovian frictional force $\gamma \dot{X}$ can be derived from the Rayleigh dissipation function, $R = \alpha \int dx \, \dot{\mathbf{n}}^2 / 2$ within the Lagrangian formalism, where $\alpha$ is the dimensionless Gilbert damping coefficient.~\cite{ThielePRL1973, TretiakovPRL2008} The stochastic force has a zero average, $\langle \nu(t) \rangle = 0$, and a white noise correlation,
\begin{equation}
\langle \nu(\omega) \nu(\omega') \rangle = 2 \pi \gamma \frac{\hbar \omega}{\tanh (\hbar \omega / 2 T)} \delta (\omega + \omega') \, ,
\end{equation}
as dictated by the quantum fluctuation-dissipation theorem.~\cite{CallenPR1951, KuboRPP1966} For high temperatures, $T \gg \hbar \omega$, it is reduced to the classical version: $\langle \nu(t) \nu(t') \rangle = 2 \gamma T \delta (t - t')$. The Gilbert damping term $\gamma \dot{X}$ and the associated stochastic force $\nu(t)$ could be absorbed into the other terms by the following transformations: $\eta(\Delta t) \mapsto \eta(\Delta t) + 2 \gamma \delta (\Delta t)$ and $\zeta(t) \mapsto \zeta(t) + \nu(t)$, but they are retained to be distinguished from the magnon-induced effects.

\subsection{Periodic force}

Let us first consider a simple situation, where the domain wall is subjected to a periodic external force:
\begin{equation}
F(t) = F_0 \cos (\omega t) \, .
\end{equation}
Application of a periodic magnetic field along the easy axis, $\mathbf{H}(t) = H_0 \cos (\omega t) \hat{\mathbf{z}}$, gives rise to this force with the magnitude $F_0 = 2 M_s H_0$, where $M_s$ is the saturation magnetization per unit length. Then, the equation of motion gives the response of the velocity
\begin{equation}
\langle V(t) \rangle = \text{Re} \left[ \mu(\omega) F_0 e^{- i \omega t} \right] \, ,
\end{equation}
where the complex mobility $\mu(\omega)$ is given by
\begin{equation}
\label{eq:mu}
\mu(\omega) = \frac{1}{- i M \omega + \eta[\omega] + \gamma} \, ,
\end{equation}
and $\eta[\omega]$ is the Fourier transform of $\eta(t)$. Therefore, by observing the response of the domain-wall velocity $V(t)$ to an oscillating magnetic field, we can infer the non-Markovian part of the frictional force $\propto \eta[\omega]$, which can be compared with our results in Eq.~(\ref{eq:eta-2}).

\subsection{Harmonic potential well}
\label{sec:well}

Let us now consider the dynamics of the domain wall trapped in a harmonic potential well, which is described by an external force $F = - k X$ with a positive constant $k$. The consideration of this case is motivated by an experimental work by \textcite{SaitohNature2004}, in which the mass of a domain wall trapped in an engineered potential well has been obtained from its dynamic response to an oscillating electric current. The dynamics of the domain wall at macroscopic time scales, $t \gg \tau$, is governed by the low-frequency part of the response kernel, $\eta[\omega]$ with $\omega \tau \ll 1$. In the limit of zero frequency, $\omega \tau \rightarrow 0$, the equation of motion~(\ref{eq:eom}) becomes local in time after replacing $\eta[\omega]$ by its approximation [Eq.~(\ref{eq:eta-approx})]:
\begin{equation}
M \ddot{X} - M \tau_\text{AL} \dddot{X} + \gamma \dot{X} = - k X + f(t) \, ,
\end{equation}
where $\tau_\text{AL} \equiv \eta_0 \tau^2 / M$ and $f(t) \equiv \zeta(t) + \nu(t)$ represents the sum of the two stochastic Langevin forces. The second term, which is third order in time derivative, is known as the Abraham-Lorentz force,~\cite{AbrahamAP1903} which gives rise to the causality paradox as follows. The response function $\chi(\omega)$ in $X(\omega) = \chi(\omega) f(\omega)$ is given by
\begin{equation}
\chi (\omega) = \frac{1}{M (\omega_t^2 - \omega^2 -  i \tau_\text{AL} \omega^3) - i \gamma \omega} \, ,
\end{equation}
where $\omega_t \equiv \sqrt{k / M}$ is the undamped frequency of the domain-wall oscillation. This response function has a pole in the upper half-part of the complex plane of $\omega$. For example, the pole is at $\omega_\text{AL} \approx i \tau_\text{AL}^{-1}$ for sufficiently small $\omega_t \ll \tau_\text{AL}^{-1}$ and $\gamma \ll M \tau_\text{AL}^{-1}$. The pole in the upper half-plane implies the existence of exponentially diverging solutions, thereby causing the famous paradox of the Abraham-Lorentz force. This paradox is an artifact of the approximation taking the zero-frequency limit, $\omega \tau \rightarrow 0$. Indeed, the location of the pole is where the zero-frequency limit is not valid, $|\omega_\text{AL} \tau| = (8 \pi / \kappa) (s \lambda / \hbar) \exp(\epsilon_0 / T) \gg 1$.

This problem can be regularized by treating the non-Markovian friction as a perturbation to the zeroth-order equation of motion, $M \ddot{X} = F$, by following Jackson.~\cite{Jackson} The regularization is executed by modifying the force term as follows:
\begin{equation}
M \ddot{X} = F + \tau_\text{AL} \frac{d F}{d t} = F + \tau_\text{AL} \left[ \frac{\partial F}{\partial t} + \dot{X} \frac{dF}{dX} \right] \, ,
\end{equation}
which does not yield runaway solutions or acausal behavior. According to \textcite{Jackson}, it is a sensible alternative to the Abraham-Lorentz equation for small radiative effects. Then, the equations of motion for $F = - k X$ is given by
\begin{equation}
M \ddot{X} + (\tau_\text{AL} k + \gamma) \dot{X} + k X = f(t) \, .
\end{equation}
The corresponding response function is 
\begin{equation}
\chi (\omega) = \frac{1}{M (\omega_t^2 - \omega^2) - i (\tau_\text{AL} M \omega_t^2 + \gamma) \omega} \, ,
\end{equation}
whose poles are in the lower half-plane. Note that the effective frictional force is sensitive to the trap frequency $\omega_t$, and, for that reason, can be distinguished from the extrinsic Ohmic friction force $\propto \gamma$. See \textcite{EfimkinPRL2016} for an analogous discussion on the friction of bright solitons in superfluids.

\subsection{Experimental considerations}

The magnon-induced effects can be inferred via the complex mobility $\mu(\omega)$ [Eq.~(\ref{eq:mu})] of a domain wall subjected to an oscillating magnetic field. The magnon-induced non-Markovian friction is comparable to the Markovian friction stemming from the Gilbert damping when $J(\omega) \sim \gamma = 2 \alpha s / \lambda$. From the analytical expression of $J(\omega)$ in Eq.~(\ref{eq:eta-approx}) and its numerical calculation in Fig.~\ref{fig:fig2}(b), this criteria can be cast as
\begin{equation}
(\omega \tau)^2 e^{ - \epsilon_0 / T} \sim \frac{\alpha s \lambda}{\hbar} \, .
\end{equation}
Note that the right-hand side includes the factor $s \lambda / \hbar$, which is the total spin contained within the domain-wall width in units of $\hbar$. 

To obtain numerical estimates for the temperature $T$ and the driving frequency $\omega$ that are suitable for probing the non-Markovian friction, let us take the parameters of a long strip of yttrium iron garnet (YIG) with thickness $t = 5$ nm and width $w = 20$ nm: $\alpha = 10^{-4}$, $s = 10^{-22}$ J$\cdot$s/m, $A = 5 \times 10^{-28}$ J$\cdot$m, $K_e = 9 \times 10^{-13}$ J/m, and $K_h = 3 \times 10^{-12}$ J/m,~\cite{BhagatPSS1973, *TakeiPRL2014} where the shape anisotropy~\cite{OsbornPR1945} induced by the dipolar interaction is taken into account.~\footnote{$K_h = 4 \pi M_s^2 w / (t + w)$ and $K_e = 4 \pi M_s^2 t / (t + w)$ in Gaussian units, where $M_s = \gamma s / t w$ is the saturation magnetization and $\gamma$ is the gyromagnetic ratio. The demagnetizing factors are for infinitely long wire with an elliptical cross section characterized by two semi axes $t$ and $w$.} This set of parameters yield the domain-wall width $\lambda \sim 20$ nm, the spin-wave gap $\epsilon_0 \sim 80$ mK, the characteristic time scale $\tau \sim 100$ ps, and the aspect ratio $\kappa = 4$. The domain wall contains enough spin, $s \lambda / \hbar \sim 2 \times 10^4$, to justify the assumption that the domain wall is much heavier than magnons (see Appendix~\ref{app:mass} for the relevant discussion). Based on these estimates, the magnon-induced non-Markovian friction and the Gilbert-damping-induced Markovian friction will be comparable when the temperature and the driving frequency are of the order of the spin-wave gap, $T \sim 50$ mK and $\omega \sim 1$ GHz. For temperatures and frequencies higher than the spin-wave gap, the magnon-induced friction is expected to dominate the Gilbert-damping-induced Markovian friction as long as the temperature is much below than the ordering temperature, $T \ll T_c$, so that our assumption of dilute magnons holds. The investigation of the frequency dependence of the complex mobility $\mu(\omega)$ under the aforementioned setting will allow us to probe the magnon-induced friction.

\section{Summary and discussion}
\label{sec:discussion}

We have derived the deterministic frictional force and the stochastic Langevin force on a domain wall in a one-dimensional ferromagnet, which is induced by its coupling to the quantum magnon bath, within the two different approaches: the time-dependent perturbation theory in quantum mechanics and the Keldysh formalism. The derivation has been facilitated by the availability of the exact solutions of spin waves on top of a domain wall. We have studied the effects of the non-Markovian friction on the dynamic response of a domain wall and have discussed a possible experimental setup to probe it.

Next, let us compare our works with those of the existing literature. First, \textcite{StampPRL1991} in 1991 and \textcite{ChudnovskyPRB1992} in 1992 have studied the problem of the magnon-induced diffusion of a ferromagnetic domain wall. Their interaction term between magnons and the domain wall comes from the Hamiltonian $H$ [Eq.~(\ref{eq:H})], in which magnons experience the velocity of the domain wall via the effective potential well, i.e., $\propto \sech^2 (x - V t)$. The resultant dissipation kernel associated with two-magnon scattering~\footnote{In three-dimensional ferromagnets, where a domain wall forms a two-dimensional plane, three-magnon scattering that involves two ``bulk'' continuum magnons and one ``wall'' magnon confined to the wall plane can give rise to an Ohmic dissipation process.~\cite{StampPRL1991, ChudnovskyPRB1992} However, this process is absent in strictly one-dimensional ferromagnets, where a domain wall is a point, and thus, for sufficiently thin ferromagnets, it can be neglected as noted in the footnote 45 of \textcite{BraunPRB1996}.} is of fourth order in the domain-wall velocity, $\propto V^4$.~\cite{ChudnovskyPRB1992} On the other hand, our interaction term [Eq.~(\ref{eq:L-coupling})] is from the spin Berry phase and the corresponding energy dissipation is of zeroth order in the domain-wall velocity, which should dominate the aforementioned force for low-energy dynamics. Second, \textcite{BraunPRB1996} studied a similar problem within the Matsubara formalism~\cite{MatsubaraPTP1955} by mapping the model for a ferromagnet to the sine-Gordon model in the limit of strong hard-axis anisotropy, $K_h \gg K_e$. Note that this assumption is not made in our work. See Appendix~\ref{app:Braun-limit} for a brief discussion on the limit of strong hard-axis anisotropy. In addition, they focused on elastic scattering of magnons off the domain wall by disregarding the off-diagonal components in the scattering matrix $\pi_{k k'}$, and, for this reason, the obtained spectral function is finite at frequencies above the spin-wave gap. In our work, we have included the effects of inelastic scattering of magnons off the domain wall and have thereby obtained a gapless spectral function, which can be expected to govern the low-energy dynamics of the domain wall.

We have made a few approximations in the paper. First, we have treated a ferromagnetic wire as a strictly one-dimensional system by assuming uniform spin configurations across the cross section, which is valid for sufficiently thin wires or low temperatures. Secondly, we have assumed that the domain-wall mass is much larger than the magnon mass, which is valid for sufficiently long domain walls, i.e., $\lambda \gg \hbar / s$. Thirdly, we have focused on only two-magnon scattering process with a domain wall by assuming dilute magnon densities, which are valid for temperatures much smaller than the ordering temperature, $T \ll T_c$. Fourthly, the Gilbert damping, which is induced by the coupling between the magnetization and the other external degrees of freedom such as lattice, has been assumed to form a featureless background for the magnetization dynamics by being local in space and time in this paper, although it can be nonlocal in both.~\cite{BrataasPRL2008, *BaratiPRB2014}

\section{Outlook}
\label{sec:outlook}

The structure of our theory for the magnon-induced friction of a domain wall in magnets is analogous to that for the quasiparticle-induced friction of a soliton in superfluids,~\cite{EfimkinPRL2016} which allows us to connect two different states of matter: magnets and superfluids. The link between them is the shared two-component description of their dynamics, which consists of the coherent order-parameter dynamics and the incoherent small-amplitude fluctuations. For both a magnetic domain wall and a superfluid soliton, the induced frictional force is a macroscopic manifestation of the interaction between the two components. This link between magnets and superfluids can be also found in the two-fluid theory for spin superfluids in easy-plane magnets,~\cite{FlebusPRL2016} which has been recently developed motivated by the two-fluid theory for superfluid helium-4.~\cite{TiszaNature1938, *LandauPR1941} We envision that multi-component description of the dynamics of ordered media may serve as a versatile link between different subfields of physics for nonequilibrium phenomena.

We have developed the theory of the domain-wall friction induced by the magnon bath in equilibrium based on the Keldysh formalism. Since the Keldysh technique is applicable to systems away from equilibrium, our theory can be a good starting point to study similar problems further out of equilibrium, e.g., the dynamics of a domain wall in the presence of a temperature gradient.~\cite{HinzkePRL2011, *YanPRL2011, *KovalevEPL2012, *SchlickeiserPRL2014, *KimPRB2015, *KimPRB2015-2} In addition, from the generalized Langevin equation that we obtained in this work, the generalized Fokker-Planck equation can be derived to study the Brownian motion of domain walls, which would exhibit anomalous behavior associated with the non-Markovian nature of the viscous and stochastic forces.~\cite{AdelmanJCP1976, *AdelmanMP1977} 

The approach taken in this paper to study the magnon-induced friction of a ferromagnetic domain wall can be also applied to the following problems. First, it has been recently shown that a magnetic domain wall in an elastic magnetic wire can be driven by the phonon current.~\cite{KimPRL2016-3} In that work, the scattering of phonons off of a domain wall has been already worked out analytically, starting from which one may develop a phonon version of our magnonic theory of the friction of a domain wall. Secondly, an analogous theory can be developed for the quantum friction of an antiferromagnetic domain wall within the Keldysh formalism staring from our earlier work on magnon-induced antiferromagnetic domain-wall motion.~\cite{KimPRB2014} This study can complement the previous results obtained by \textcite{IvanovJPCM1993} within the time-dependent perturbation theory. The Lagrangian of antiferromagnets is invariant under the Lorentz-like transformations,~\cite{HaldanePRL1983, *BaryakhtarJETP1983} which may facilitate the development of the theory. Similarly to the case of ferromagnets, magnons pass through a domain wall without any reflection in antiferromagnets,~\cite{IvanovJPCM1993} and thus the magnon-induced friction is expected to be non-Markovian despite the existence of the Lorentz-like invariance. Lastly, a system considered in this work is a one-dimensional ferromagnet with  easy-axis anisotropy, $K_e > 0$, which has two different ground states and thus can harbor a topological soliton---domain wall---interpolating them. However, in the case of an easy-plane ferromagnet without easy-axis anisotropy, $K_e = 0$, ground states are continuously degenerate and thus there is no domain wall. Instead, the system is known to support the other types of solitonic nonlinear excitations,~\cite{KosevichJETP1983, *KosevichPR1990} whose magnon-induced friction can be investigated within the same formalism used in this work. In addition, the analogy between easy-plane magnets and superfluids~\cite{HalperinPR1969, *SoninJETP1978} may allow us to identify the magnetic counterparts of the known results for friction of superfluid solitons.~\cite{EfimkinPRL2016, HurstPRA2017}

\begin{acknowledgments}
This work was supported by the Army Research Office under Contract No. W911NF-14-1-0016 (S.K.K. and Y.T.). Work at JHU was supported by the US Department of Energy, Office of Basic Energy Sciences, Division of Materials Sciences and Engineering under Award DE-FG02-08ER46544 (O.T.). V.M.G. acknowledges support from the DOE-BES (DESC0001911) and the Simons Foundation.
\end{acknowledgments}

\appendix

\section{The total momentum of a domain wall with spin waves}
\label{app:conserved}

We derive the total momentum of a domain wall with spin waves on top of it by using the collective-coordinate approach,~\cite{TretiakovPRL2008} which allows us to obtain the conserved momentum that is independent of a gauge choice for the spin Berry phase.~\cite{TchernyshyovAP2015} The conserved momentum of a domain wall can be obtained by the following expression:
\begin{equation}
P[\mathbf{q}(t)] = - \int^{\mathbf{q}(t)}_{\mathbf{q}_0} dq'_i \, G_{X q'_i} \, ,
\end{equation}
where $X$ is the collective coordinate for the domain-wall position, $\{ q_i \}$ represents the all the other collective coordinates including the azimuthal angle $\Phi$ and spin-wave modes, $\mathbf{q}_0$ is an arbitrary initial value of the collective coordinates, $G_{X q_i}$ is the gyrotropic tensor given by 
\begin{eqnarray}
G_{X q_i} &=& - \int dx \, \mathbf{n} \cdot \left( \frac{\partial \mathbf{n}}{\partial X} \times \frac{\partial \mathbf{n}}{\partial q_i} \right) \\
&=& \int dx \, \mathbf{n} \cdot \left( \frac{\partial \mathbf{n}}{\partial x} \times \frac{\partial \mathbf{n}}{\partial q_i} \right) \, .
\end{eqnarray}
The resultant conserved momentum is given by
\begin{equation}
\begin{split}
P[\Phi(t), \tilde{\mathbf{q}}(t)] = & - \left[ \int dx \, \mathbf{n} \cdot \left( \frac{\partial \mathbf{n}}{\partial x} \times \frac{\partial \mathbf{n}}{\partial \Phi} \right) \right] \Phi(t) \\
& - \int dx d\tilde{q}'_i \, \mathbf{n} \cdot \left( \frac{\partial \mathbf{n}}{\partial x} \times \frac{\partial \mathbf{n}}{\partial \tilde{q}'_i} \right) \, ,
\end{split}
\end{equation}
where $\{ \tilde{q}_i \}$ are the collective coordinates for the continuum modes besides the two localized modes described by $X$ and $\Phi$. Here, the first term is the contribution from a domain wall, whereas the second term that we denote by $P_m$ is the contribution from spin waves. By using the linear expansion, $\mathbf{n}(x, t) \approx \mathbf{n}_0 (x) + \delta \theta [x, \tilde{\mathbf{q}}(t)] \partial_\theta \mathbf{n}_0 (x) + \delta \phi [x, \tilde{\mathbf{q}}(t)] \partial_\phi \mathbf{n}_0 (x)$ and performing an integration over $\tilde{\mathbf{q}}$, we obtain the first term, $2 \Phi$, and the second term as follows:
\begin{widetext}
\begin{eqnarray}
P_m[\tilde{\mathbf{q}}] &=& \int dx \int^{\tilde{\mathbf{q}}}_{\tilde{\mathbf{q}}_0} d\tilde{q}_i' \, \sin \theta_0 (x) \left( \frac{\partial \delta \theta [x, \tilde{\mathbf{q}}']}{\partial \tilde{q}'_i} \, \partial_x \delta \phi [x, \tilde{\mathbf{q}}'] - \frac{\partial \delta \phi [x, \tilde{\mathbf{q}}']}{\partial \tilde{q}'_i} \, \partial_x \delta \theta [x, \tilde{\mathbf{q}}'] \right) \, , \\
&=& \int dx \, \sin \theta_0 (x) \int^{\tilde{\mathbf{q}}}_{\tilde{\mathbf{q}}_0} d\tilde{q}_i' \, \left( \frac{\partial \delta \theta [x, \tilde{\mathbf{q}}']}{\partial \tilde{q}'_i} \, \partial_x \delta \phi [x, \tilde{\mathbf{q}}'] + \frac{\partial \{ \partial_x \delta \phi [x, \tilde{\mathbf{q}}'] \}}{\partial \tilde{q}'_i} \, \delta \theta [x, \tilde{\mathbf{q}}'] \right) \, , \\
&=& \int dx \, \sin \theta_0 \, \delta \theta[x, \tilde{\mathbf{q}}] \partial_x \delta \phi[x, \tilde{\mathbf{q}}] \, ,
\end{eqnarray}
\end{widetext}
to quadratic order in $\Phi, \delta \theta$, and $\delta \phi$. Here, we drop the contribution from the integrand $(\partial_x \sin \theta_0 (x)) \, \delta \theta \delta \phi$ that vanishes on the timescale longer than the inverse of the gap frequency $\omega_0^{-1}$, and the boundary term ($\propto \delta \theta[x, \tilde{\mathbf{q}}_0] \partial_x \delta \phi[x, \tilde{\mathbf{q}}_0]$) from the arbitrary initial collective-coordinate value $\tilde{\mathbf{q}}_0$. In conjunction with the spin-wave solutions in Eqs.~(\ref{eq:n1}, \ref{eq:n2}), this conserved momentum leads to our results in the main text, $P_m$~(\ref{eq:Pm}), $\pi_{k k'}$~(\ref{eq:pi}), and $P$~(\ref{eq:P}).

\section{Mass of a domain wall versus mass of a magnon}
\label{app:mass}

We compare the mass of a domain wall and that of a magnon. For more transparent discussions, we use the physical units instead of the natural units [Eq.~(\ref{eq:unit})] in this section. The mass of a domain wall is given by 
\begin{equation}
M = \frac{2 \epsilon}{\kappa c^2} \, ,
\end{equation}
where $2 \epsilon$ and $c = \lambda / \tau$ are its rest energy and characteristic velocity, respectively. From the low-energy limit of the dispersion of a magnon [Eq.~(\ref{eq:dispersion})], 
\begin{equation}
\epsilon_k = \frac{\hbar}{\tau} \left[ 1 + \kappa + \frac{(2 + \kappa) (\lambda k)^2}{2} \right] + \text{O}(k^3) \, ,
\end{equation}
we can obtain the mass of a magnon:
\begin{equation}
\label{eq:m}
m = \frac{ \hbar \tau \sqrt{1 + \kappa}}{\lambda^2 (2 + \kappa)} \, .
\end{equation}
Their ratio is given by
\begin{equation}
\frac{M}{m} = \frac{4 + 2 \kappa}{\kappa \sqrt{1 + \kappa}} \frac{s \lambda}{\hbar} \, .
\end{equation}
Note that $s \lambda$ is the total spin inside the domain wall. When this domain-wall spin is much larger than the spin $\hbar$ of a magnon, the mass of a domain wall is much heavier than the mass of a magnon.

\section{The derivation of the frictional force based on momentum conservation}
\label{app:momentum}

Here, we invoke the conservation of the total linear momentum to obtain the magnon-induced frictional force on a domain wall within the time dependent perturbation theory, instead of the conservation of the total energy used in the main text. The force on the domain wall is the rate of the momentum transfer from the magnon bath to the domain wall, and it is given by
\begin{eqnarray}
F(t) &=& \int \frac{dk dk'}{(2 \pi)^2} R_{k k'} (t) \hbar (k - k') f_k \\
&=& \frac{1}{2} \int \frac{dk dk'}{(2 \pi)^2} R_{k k'} (t) \hbar (k - k') (f_k - f_{k'}) \, .
\end{eqnarray}
By using Eq.~(\ref{eq:R}) for $R_{k k'}$, we obtain
\begin{equation}
\label{eq:F}
\begin{split}
F(t) = \frac{1}{\hbar^2} \int_0^t dt' \int \frac{dk dk'}{(2 \pi)^2} |\pi_{k k'}|^2 (f_k - f_{k'}) V(t') \\
\times \cos \left[ \frac{(\epsilon_{k'} - \epsilon_k) (t - t')}{\hbar} \right] \hbar (k - k') V(t) \, .
\end{split}
\end{equation}
Here, the last factor $\hbar (k - k') V$ is the opposite of the change of the domain-wall energy by assuming instant interaction:
\begin{equation}
\Delta E_\text{dw} = (\Delta P_\text{dw}) V \, .
\end{equation}
From the energy conservation, this should be equivalent to the change of the magnon energy: $\hbar (k - k') V = \epsilon_{k'} - \epsilon_k$. Then, the above Eq.~(\ref{eq:F}) yields the same result for $\eta(t)$ [Eq.~(\ref{eq:eta-2})] that has been obtained by invoking the energy conservation alone in the main text.

\section{The limit of strong hard-axis anisotropy}
\label{app:Braun-limit}

Here, we discuss the limit of strong hard-axis anisotropy, $\kappa \gg 1$, in order to compare our results with those obtained by Braun and Loss.~\cite{BraunPRB1996} First, let us begin with analysis of spin-wave modes on a static domain wall. In this limit, the Hamiltonian density for $\mathcal{H}_2$ [Eq.~(\ref{eq:H2})] for $\delta n_2$ can be approximated by the hard-axis anisotropy constant, $\mathcal{H}_2 \approx \kappa$, which allows us to treat $\delta n_2$ as a slave variable of $\delta n_1$ and integrate it out. By using the equations of motion, $\delta n_2 \approx \delta \dot{n}_1 / \kappa$, and representing $\delta n_1$ by the small-angle field $\varphi$ by following Braun and Loss,~\cite{BraunPRB1996} we can obtain the Lagrangian in terms of $\varphi$ only:
\begin{equation}
L_2 \approx \int dx \left[ \dot{\varphi}^2 / 2 \kappa - \varphi \mathcal{H}_1 \varphi / 2 \right] \, .
\end{equation}
This Lagrangian corresponds to Eq.~(5.11) of Braun and Loss.~\cite{BraunPRB1996} The solutions are given by Eq.~(\ref{eq:n1}) with the ellipticity factor omitted and the dispersion is given by $\omega_k \approx \sqrt{\kappa (1 + k^2)}$.

Let us now consider a moving domain wall. To linear order in $\dot{X}$ and the quadratic order in $\delta n_1, \delta n_2$, and $\Phi$, the interaction term between a domain wall and spin waves is given by
\begin{equation}
\begin{split}
L_\text{int} = & \dot{X} \int dx (\delta n_1 \partial_x \delta n_2) \, \\
\approx & \dot{X} \int dx (\varphi \partial_x \partial_t \varphi) / \kappa \, ,
\end{split}
\end{equation}
where $\delta n_2$ is replaced by $\delta \dot{n}_1 / \kappa$ by using the equations of motion on the second line. The right-hand side on the first line gives rise to our interaction term in Eq.~(\ref{eq:L-coupling}), which is valid for an arbitrary value of $\kappa$. The approximated one on the second line, which pertains to the limit $\kappa \gg 1$, is the first term in Eq.~(5.12) of Braun and Loss.~\cite{BraunPRB1996}

\section{Details of the Keldysh calculation}
\label{app:Keldysh}

In this section, we provide the details of the Keldysh calculation, for which we closely follow the approach taken by \textcite{EfimkinPRL2016}. After integrating out the magnon fields $\xi$ from the Keldysh action [Eq.~(\ref{eq:SK})] in the one-loop approximation, we obtain the following effective action:
\begin{widetext}
\begin{equation}
S^K_\text{eff} = 2 M \int dt \dot{X}_c \dot{X}_q + \int dt dt' \left[ \dot{X}_c (t) \Pi_{cq} (t, t') \dot{X}_q (t') + \dot{X}_q (t) \Pi_{qc} (t, t') \dot{X}_c (t') + \dot{X}_q (t) \Pi_{qq} (t, t') \dot{X}_q (t') \right] \, ,
\end{equation} 
where
\begin{eqnarray}
\Pi_{cq} (t, t') &=& \Pi_{qc} (t', t) = - \frac{i}{\hbar} \int \frac{dk dk'}{(2 \pi)^2} |\pi_{kk'}|^2 \left[ G^A_{k'} (t, t') G^K_{k} (t', t) + G^K_{k'} (t, t') G^R_{k} (t', t) \right] \, , \\
\Pi_{qc} (t, t') &=& \Pi_{cq} (t', t) = - \frac{i}{\hbar} \int \frac{dk dk'}{(2 \pi)^2} |\pi_{k k'}|^2 \left[ G^K_{k'} (t, t') G^A_{k} (t', t) + G^R_{k'} (t, t') G^K_{k} (t', t) \right] \, , \\
\Pi_{qq} (t, t') &=& - \frac{i}{\hbar} \int \frac{dk dk'}{(2 \pi)^2} |\pi_{kk'}|^2 \left[ G^K_{k'} (t, t') G^K_{k} (t', t) + G^R_{k'} (t, t') G^A_{k} (t', t) + G^A_{k'} (t, t') G^R_{k} (t', t) \right] \, .
\end{eqnarray}
The absence of $\Pi_{cc}$ is required by the causality. Note that, $\Pi_{qc} (t, t') = \Pi_{qc} (t - t')$ and $\Pi_{qq} (t, t') = \Pi_{qq} (t - t')$. Using the explicit forms of Green functions [Eq.~(\ref{eq:G})], we obtain
\begin{eqnarray}
\Pi_{qc} (\Delta t) &=& - \frac{2 \Theta(\Delta t)}{\hbar} \int \frac{dk dk'}{(2 \pi)^2} |\pi_{k k'}|^2 \sin \left[ \frac{(\epsilon_{k'} - \epsilon_k) \Delta t}{\hbar} \right] (f_{k'} - f_k) \, , \\
\Pi_{qq} (\Delta t) &=&  \frac{2 i}{\hbar} \int \frac{dk dk'}{(2 \pi)^2} |\pi_{k k'}|^2 \cos \left[ \frac{(\epsilon_{k'} - \epsilon_k) \Delta t}{\hbar} \right] (f_{k'} + f_k + 2 f_{k'} f_k) \, .
\end{eqnarray}

By performing the integration by parts on the action above, we get
\begin{equation}
S^K_\text{eff} = \int dt \left[ 2 X_q (t) ( - M \ddot{X}_c ) - \int dt' \, 2 X_q (t) \eta(t - t') \dot{X}_c (t') \right] + \frac{i}{\hbar} \int dt dt' 2 X_q (t) C_s (t - t') X_q (t') \, ,
\end{equation}
where $\eta(t - t') = \partial_t \Pi_{qc} (t - t') / 2 \propto \Theta(t - t')$ and $C_s (t - t') = \hbar \partial_t \partial_{t'} \Pi_{qq} (t - t') / 4 i$. By performing the Hubbard-Stratonovich transformation for the last term by introducing an auxiliary field $\zeta$, we obtain
\begin{equation}
S^K_\text{eff} = \int dt \left\{ 2 X_q (t) \left[ - M \ddot{X}_c  - \int_0^t dt' \eta(t - t') \dot{X}_c (t') + \xi(t) \right] \right\} + \frac{i \hbar}{2} \int dt \int dt' \xi (t) C_s^{-1} (t - t') \xi (t') \, .
\end{equation}
The saddle-point solution to this action gives the quasiclassical equation of motion for the coordinate $X$:
\begin{equation}
M \ddot{X} + \int_0^t dt' \eta(t - t') \dot{X}(t') = F + \xi(t) \, 
\end{equation}
where $\xi$ satisfies
\begin{equation}
\langle \xi(t) \xi(t') \rangle = C_s (t - t') \, .
\end{equation}

The explicit expressions for $\eta(t - t')$ and $C_s(t - t')$ are given by
\begin{eqnarray}
\eta(t - t') &=& \eta(\Delta t) = \frac{\Theta(\Delta t)}{\hbar^2} \int \frac{dk dk'}{(2 \pi)^2} |\pi_{k k'}|^2 (\epsilon_{k'} - \epsilon_k) (f_{k} - f_{k'}) \cos \left[ \frac{(\epsilon_{k'} - \epsilon_k) \Delta t}{\hbar} \right] \, , \\
C_s (t- t') &=& C_s (\Delta t) = \frac{1}{2 \hbar^2} \int \frac{dk dk'}{(2 \pi)^2} |\pi_{k k'}|^2 (\epsilon_{k'} - \epsilon_k)^2 \cos \left[ \frac{(\epsilon_{k'} - \epsilon_k) \Delta t}{\hbar} \right] \coth \left[ \frac{\hbar (\epsilon_{k'} - \epsilon_k)}{2 T} \right] \, .
\end{eqnarray}
In the classical limit, where the temperature is much higher than all the energy scales of the system, $T \rightarrow \infty$, we obtain
\begin{equation}
C_s (t) = T \eta \left( |t| \right) \, ,
\end{equation}
which is a manifestation of the classical fluctuation-dissipation theorem.~\cite{FariasPRE2009, *RuckriegelPRL2015}
\end{widetext}

\bibliography{/Users/evol/Dropbox/School/Research/master}

\end{document}